\documentclass[twocolumn,showpacs,prb,preprintnumbers,amsmath,amssymb]{revtex4}
\usepackage{graphicx}
\usepackage{dcolumn}
\usepackage{multirow}
\usepackage{epsfig}
\begin{document}
\title{Phonon Control of Magnetic Relaxation in the Pyrochlore Slab Compounds SCGO and BSZCGO}
\author{M. Zbiri}
\email[Corresponding author. Electronic address:~]{zbiri@ill.fr}
\affiliation{Institut Max von Laue-Paul Langevin, 6 rue Jules Horowitz, BP 156, 38042 Grenoble Cedex 9, France}
\author{H. Mutka}
\affiliation{Institut Max von Laue-Paul Langevin, 6 rue Jules Horowitz, BP 156, 38042 Grenoble Cedex 9, France}
\author{M. R. Johnson}
\affiliation{Institut Max von Laue-Paul Langevin, 6 rue Jules Horowitz, BP 156, 38042 Grenoble Cedex 9, France}
\author{H. Schober} 
\affiliation{Institut Max von Laue-Paul Langevin, 6 rue Jules Horowitz, BP 156, 38042 Grenoble Cedex 9, France}
\author{C. Payen}
\affiliation{Institut des Mat\'eriaux Jean Rouxel, Universit\'e de Nantes - CNRS, B.P. 32229, 44322 Nantes Cedex 3 - France}
\begin{abstract}
We are interested in  the  phonon response in the frustrated magnets SrCr$_{9x}$Ga$_{12-9x}$O$_{19}$ (SCGO) and Ba$_{2}$Sn$_{2}$ZnCr$_{7x}$Ga$_{10-7x}$O$_{22}$ (BSZCGO). The motivation of the study is the 
recently discovered, phonon-driven, magnetic relaxation in the SCGO compound  [Mutka et al. PRL {\bf 97} 047203 (2006)] pointing out the importance of a low-energy ($\hbar\omega\sim$7~meV) phonon mode. 
In neutron scattering experiments on these compounds  the phonon signal is partly masked by the magnetic signal from the Cr moments and we have therefore examined in detail  the non-magnetic 
isostructural counterparts SrGa$_{12}$O$_{19}$ (SGO) and Ba$_{2}$Sn$_{2}$ZnGa$_{10}$O$_{22}$ (BSZGO). Our  {\it ab-initio} lattice dynamics calculations on SGO reveal a peak in the vibrational density of 
states matching with the neutron observations on SGO and SCGO.  A strong contribution in the vibrational density of states comes from the partial contribution of the Ga atoms on the 2b and 12k sites, 
involving modes at the M--point of the hexagonal system.  These modes comprise dynamics of the kagom\'e planes of the pyrochlore slab magnetic sub-lattice, 12k sites, and therefore can drive magnetic 
relaxation via spin-phonon coupling. Both BSZCGO and BSZGO show a similar low-energy Raman peak but no corresponding peak in the neutron determined density of states of BSZGO is seen. However,  a strong 
non-Debye enhancement of low-energy phonon response is observed. We attribute this particular feature to the Zn/Ga disorder on the 2$d$ -site, already evoked earlier to affect the magnetic properties of 
BSZCGO. We propose that this disorder-induced phonon response explains the absence  of  a characteristic energy scale and the much faster magnetic relaxation observed  in BSZCGO.
\end{abstract}
\maketitle
\section{Introduction}
The interplay between structural, electronic and dynamic degrees of freedom in geometrically  frustrated magnetic materials has the consequence of creating highly  degenerate ground states, which have 
generated considerable interest~\cite{ram01, HFM2003, HFM2006}. Short-range correlations in frustrated magnets lead to the formation of weakly coupled, fluctuating clusters  and consequently a 
macroscopic, collective degeneracy can prevail~\cite{moes98a, moes98b, ram01,moes06}.\\ 
At low temperature, frustrated magnets are expected to be sensitive to weak perturbations, which can raise the ground-state degeneracy, produce a hierarchy of closely space energy levels and allow the 
possibility of slow dynamics at low temperature between the corresponding states. When the associated energy scales of these states fall well below that of 
magnetic interactions~\cite{moes01,moes06}, various perturbations including static~\cite{tchern02b} or dynamic~\cite{sush05, fennie06} lattice effects  may play a key role due to coupling between lattice and magnetic degrees of freedom. A recent 
neutron spin--echo (NSE) examination of  the pyrochlore slab antiferromagnet (kagome bilayer) SrCr$_{9x}$Ga$_{12-9x}$O$_{19}$ (SCGO)~\cite{mut06} suggested that phonons affect the slow relaxation of this 
highly frustrated spin system. The SCGO compound has a particular ground-state in which dimer correlations within the pyrochlore slabs show a partial freezing of about 1/3 of the total possible ordered  
magnetic moment in spite of the high value of the average intraslab superexchange interaction ($\sim$ 13 meV)~\cite{limot02}. The frozen ground-state does not fit into a spin-glass picture, for example 
it has been shown that both the freezing transition temperature and the frozen moment  decrease with increased level of magnetic dilution due the Ga/Cr substitution~\cite{mond01}.\\ 
The extremely broad  relaxation seen using NSE  spectroscopy was fitted with a phenomenological stretched exponential  time--dependence to infer an  activated temperature dependence  with an activation 
energy of $E_{a}\sim$~7~meV~\cite{mut06}. A vibrational mode was observed close to  this same position by Raman spectroscopy and inelastic neutron scattering (INS), leading to the conclusion that phonons 
drive the relaxation.
\begin{table*}[hptb]
 \caption{Lattice parameters and  the relevant Ga/Cr position parameters from  \emph{ab-initio} calculations on SGO compared with experimental data on SGO and SCGO. 
\label{tab_params}}
 \begin{ruledtabular}
 \begin{tabular}{c|ccccc}
parameter & SGO (calc)  & \multicolumn{3}{c}{SGO (exp)~\cite{grae94}} & SCGO (exp)~\cite{obra88a}  \\ 
\hline
         &              & $T$=473~K & $T$=298~K & $T$=16~K & $T$=4.2~K \\
\hline
a(\AA) & 5.86 & 5.80  & 5.80  & 5.79  & 5.80 \\
c(\AA) & 22.87 & 22.86 & 22.82 & 22.78 & 22.66 \\
2a:$x,y,z$ & 0, 0, 0 & 0, 0, 0 & 0, 0, 0 & 0, 0, 0 & 0, 0, 0\footnotemark[2]\\
2b(4e):$x,y,z$\footnotemark[1]\footnotemark[3] & 0, 0, 1/4 & 0, 0, 1/4(0.258) & 0, 0, 1/4(0.257) & 0, 0, 1/4 (0.256) & 0, 0, 1/4(0.256)\\
12k:$x,z$  & 0.1678, 0.891  & 0.1684, 0.891 & 0.1683, 0.891 & 0.1682, 0.891 & 0.1681, 0.892\footnotemark[2]  \\
4f$_{iv}$:$z$\footnotemark[3] & 0.0278 & 0.0273 & 0.0273 & 0.0274 & 0.0283 \\
4f$_{vi}$:$z$  & 0.190  & 0.190 & 0.190 & 0.190 & 0.192\footnotemark[2]  \\
 \end{tabular}
 \end{ruledtabular}
 \footnotetext[1]{For refinement of experimental data it is possible to split the 2b occupation over two 4e sites~\cite{grae94,obra88a}.} 
 \footnotetext[2]{Cr occupation $\gtrsim$~90\% in least diluted samples~\cite{mond01}. The Kagome (12k) and triangular (2a) planes form the pyrochlore slab.}
 \footnotetext[3]{Ga occupation 100~\%.}
 \end{table*}
However, why phonons at 7~meV, and not other frequencies, drive magnetic relaxation is not clear and further investigation of phonons and related spin-phonon mechanisms is required. In this context we 
have studied the isostructural  non-magnetic material SrGa$_{12}$O$_{19}$ (SGO) using ab-initio lattice dynamics calculations that provide a full picture of the phonon dispersion relations, total 
and partial density of states, energies of Raman active modes and the neutron scattering cross-section. The choice of the non-magnetic counterpart SGO is motivated by experimental considerations 
(see below) and the fact that, practically, handling the chemical disorder associated with the Cr/Ga substitution in SCGO efficiently is not feasible in the lattice dynamics calculation.The calculated 
neutron scattering cross-section is used  for the evaluation of the powder averaged $Q$--dependent intensity that can be compared  with experiment. The calculations are accompanied by new experiments 
using INS and Raman spectroscopy. Experimental techniques are combined because Raman spectroscopy gives a partial view of the vibrational density of states ($\Gamma$--point only) which is complemented by 
INS that can inform on  the $Q$-dependence and characteristic energies throughout the Brillouin zone.\\
In addition to the SCGO/SGO case  we have examined the experimental situation in the related pyrochlore slab compound Ba$_{2}$Sn$_{2}$ZnCr$_{7x}$Ga$_{10-7x}$O$_{22}$ (BSZCGO) and its non-magnetic counterpart 
Ba$_{2}$Sn$_{2}$ZnGa$_{10}$O$_{22}$ (BSZGO). Ab-initio lattice dynamics calculations could not be done for BSZCGO and BSZGO due to their complex crystal structures. In BSZCGO the magnetic relaxation rate 
at low-$T$ was observed to be by some two orders of magnitude faster  than in SCGO but  without a well-defined energy scale~\cite{mut06}. Nevertheless,  both in BSZCGO and BSZGO Raman data reveal a phonon 
peak  at an energy close to the one seen in SCGO and SGO, while no peak is seen in  the INS response. As we shall see below the  absence of a characteristic energy in the relaxation of BSZCGO makes more 
sense when we consider the rather strong non-Debye enhancement of the phonon DOS that we attribute to the substitutional disorder, the 50/50 mix of Zn and Ga on the 2$d$ site in this 
compound~\cite{bonnet04}. Accordingly we conclude that the localized phonon modes associated with the substitutional disorder can also affect the magnetic interactions and induce relaxational dynamics 
of the magnetic system. These observations also point out the particularity of the frustrated magnets concerning the system dependence of the low--energy properties.\\
As for SCGO, the aim of the present work is (i) to identify from {\it ab-initio}  calculations the experimentally observed phonon modes (neutron and Raman spectroscopy), these calculations were used to 
obtain generalized neutron weighted vibrational density of states (GDOS) and the active Raman modes for SGO, (ii) understand the importance of non-$\Gamma$-point modes, and (iii) investigate the normal 
modes having an effect on the dynamics of the magnetic sites and hence the possibility to drive magnetic relaxation.\\
This paper is organized as follows: The experimental and computational details are provided in Section \ref{expmeth} and Section \ref{comp}, respectively. Section \ref{results} is dedicated to the 
presentation of the results that are discussed in Section \ref{disc} with  conclusions.
\section{Experimental Details\label{expmeth}}
Powder samples of non-magnetic SGO and BSZGO were prepared using a standard solid-state high temperature ceramic synthesis method and characterized by x-ray and neutron diffraction. As for 
the magnetic SCGO (x=0.95) and BZSCGO (x=0.97) samples, we analyzed those that were already used in our previous studies~\cite{mond01,bonnet04,mut06,mut07}.
The INS measurements were performed at the Institut Laue Langevin (ILL)~\cite{ill} (Grenoble, France), using two instruments; the cold neutron time-of-flight spectrometer IN6 and the thermal neutron 
time-of-flight spectrometer IN4. IN6 operating with an incident wavelength of $\lambda_{i}$~= 4.1 {\AA} provides very good resolution (0.2~meV FWHM)  in the lower energy transfer 
range ($|\hbar\omega|\le$~10~meV) for the anti-Stokes spectrum.  On IN4 using incident wavelengths of  $\lambda_{i}$~= 2.6 or 1.8 {\AA} one has an extended $Q$-range and this allows the Stokes 
spectrum to be measured at low temperature over a broader energy transfer range with a resolution of  0.5 or 1 meV (FWHM), respectively. Experiments were performed on both SCGO/SGO  and BSZCGO/BSZGO 
systems mainly at $T$~=~300K and also at low temperatures 2~K$<T<$12~K for SCGO/SGO. The data analysis was done using ILL software tools and the GDOS was evaluated using standard 
procedures~\cite{mut08} without applying multiphonon corrections. The experimental GDOS was normalised to  3N modes where N is the number of atoms in the formula unit. Raman measurements at $T$~=~300~K 
were performed  at the Institut des Mat\'eriaux Jean Rouxel (IMN)~\cite{imn}. 
%
\begin{figure*}
\includegraphics[angle=270,width=160mm]{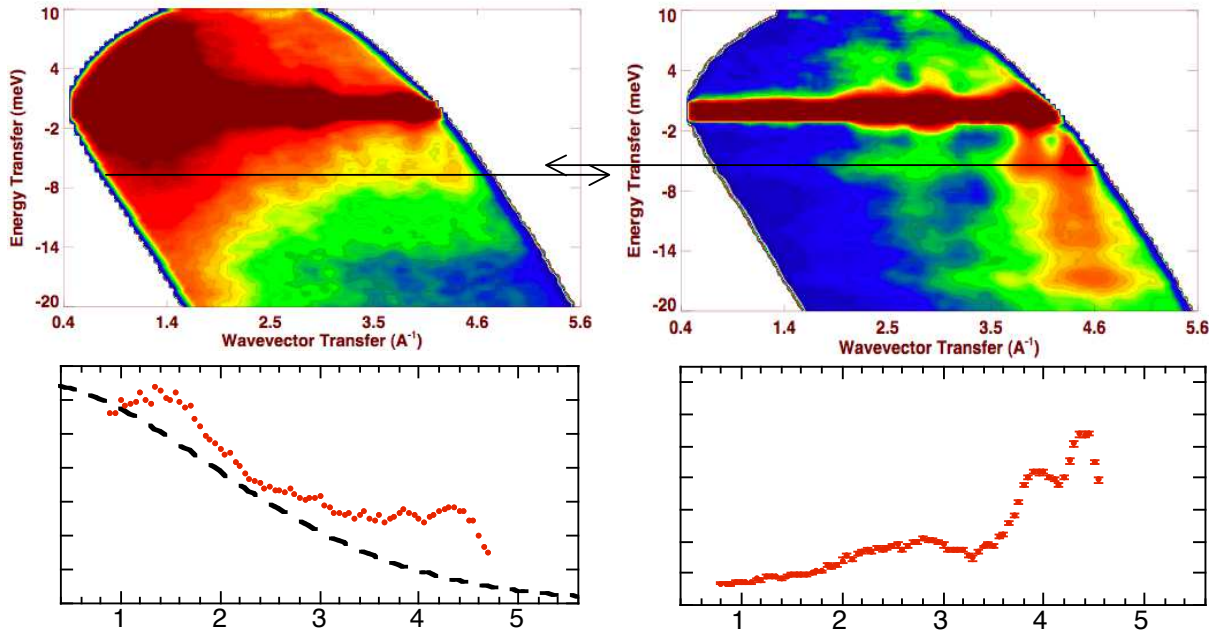}
 \caption{(Color online) Measured $S(Q,\omega)$ maps at $T$=~300~K for SCGO (top left) and  SGO (top right). Strong paramagnetic scattering  dominates the low--$Q$ response of SCGO.  We can note that a 
peaked phonon  reponse comparable to the one seen  in SGO at $\hbar\omega\approx$~-5.5~meV with $Q$--dependent intensity is observed in SCGO at the highest $Q$--range, but at a slightly shifted energy  
transfer $\hbar\omega\approx$~-7~meV, as shown by the arrows. The bottom panels show the constant energy transfer scans  at $\hbar\omega~=$~- 7~meV for SCGO (bottom left) and  at $\hbar\omega~=$~- 5.5~meV 
for SGO (bottom right). For SCGO   the magnetic form factor $Q$--dependence  is shown (dashed line) in order to indicate the trend of the magnetic scattering. The phonons in SCGO appear less structured 
and somewhat blurred due to the weaker coherent scattering power of Cr--atoms as well as due to the overlap with the  magnetic signal.}
\label{sqwexp} 
\end{figure*}
\section{Computational Details\label{comp}}
Relaxed geometries and total energies were obtained using the projector-augmented wave (PAW) formalism~\cite{paw} of the Kohn-Sham formulation of the density functional theory KSDFT~\cite{ks1,ks2} at
the generalized gradient approximation level (GGA), implemented in the Vienna {\it ab-initio} simulation package (VASP)~\cite{vasp1,vasp2}. The GGA was formulated by the Perdew-Burke-Ernzerhof 
(PBE)~\cite{pbe1,pbe2} density functional. All results are well converged with respect to {\it k}-mesh and energy cutoff for the plane wave expansion. The break conditions for the self consistent field 
(SCF) and for the ionic relaxation loops were set to 10$^{-6}$ eV and 10$^{-4}$ eV \AA$^{-1}$, respectively. Hellmann-Feynman forces following geometry optimisation were less than 10$^{-4}$ eV \AA$^{-1}$.
Full geometry optimization, including cell parameters, was carried out on the experimentally refined SGO structure~\cite{grae94} containing eleven crystallographically inequivalent atoms (5 O, 5 Ga and 1 Sr). A comparison of the \emph{ab-initio} optimized and experimental structural parameters is given in the Table~\ref{tab_params}. The space group is P6$_3$/mmc  with 2 formula-units per unit cell (64 atoms). In order to determine all force constants, the supercell approach was used for lattice dynamics calculations. An orthorhombic supercell was constructed from the relaxed structure containing 8 formula-units (256 atoms). A second partial geometry optimization (fixed lattice parameters) was performed on the supercell in order to further minimize the residual forces. Total energies and Hellmann-Feynman forces were calculated for 90 structures resulting from individual displacements of the symmetry inequivalent atoms in the supercell, along the inequivalent cartesian directions ($\pm$x, $\pm$y and $\pm$z). 192 phonon branches corresponding to the 64 atoms in the primitive cell, were extracted from subsequent calculations using the direct method~\cite{direct} as implemented in the Phonon software~\cite{phonon}.
The coherent dynamic structure factor $S(Q,\omega)$ was calculated by the numerical procedure for evaluation of powder-averaged lattice dynamics, PALD~\cite{john09, koza08}. For a list of wave 
vectors $K$ with randomly chosen directions, but lengths corresponding to the Q range of interest, the dynamical matrix is diagonalised for each wave vector and the corresponding eigenfrequencies and 
eigenvectors are used to calculated $S(Q,\omega)$ for one-phonon creation. The wave vectors, spectral frequencies and intensities are then used to construct the 2-D $S(Q,\omega)$ map, which can be 
compared with the measured map as will be shown below.
\section{Results\label{results}}
\subsection{SGO/SCGO phonon response from experiments\label{sgo}}
Figure \ref{sqwexp} shows  the $S(Q,\omega)$ intensity maps  measured at $T$~=~300~K using  the IN4 spectrometer, for both the magnetic system SCGO and its non-magnetic counterpart SGO. The strong 
intensity at low $Q$ centered in the elastic position in SCGO is due to the quasi-elastic magnetic response resulting from the fluctuating  $S=3/2$ moments of the Cr$^{3+}$ atoms. Note that 
the $Q$--dependence of the magnetic response is modulated with a main maximum at $Q$=~1.4~\AA$^{-1}$ and the paramagnetic form factor gives just an overall trend. In spite of the decay of the magnetic 
form factor with increasing $Q$, this signal dominates and swamps the phonon signal. The $S(Q,\omega)$--map of  SGO shows the phonon signal clearly visible in the absence of the magnetic response and 
it is possible to correlate the 7~meV features  at highest $Q$--range, $Q\ge$~ 3.5~\AA$^{-1}$, in SCGO with the 5.5 meV features in  non-magnetic sample SGO. As we can see phonons in SCGO are masked 
due to the presence of the magnetic signal and therefore are difficult to measure. 
\begin{figure}
\includegraphics[angle=0,width=75mm]{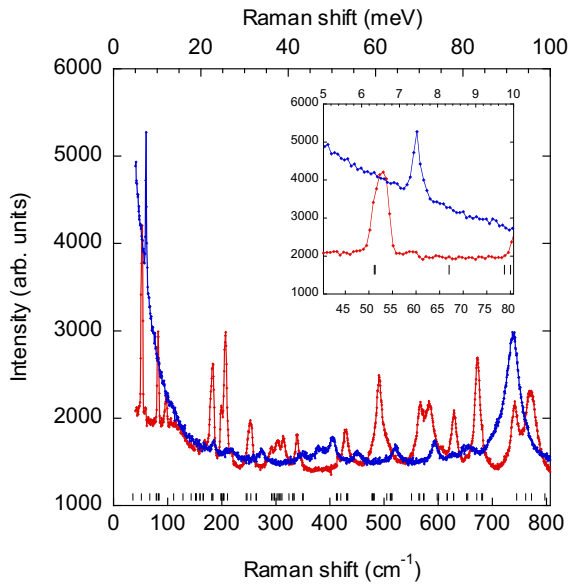}
 \caption{(Color online) Raman spectra  at $T$=~300~K for SGO (red), with calculated Raman active mode energies shown by the ticks at the bottom of the graph,  and  SCGO (blue, higher intensity at low energy). Both of these compounds have a Raman active low--energy mode in  the vicinity of the kink seen in the GDOS, compare with Fig.\ref{GDOS1}. The inset shows in detail the low--energy range highlighting the good match with the position calculated for SGO.}
 \label{raman1} 
\end{figure}
\begin{figure}
\includegraphics[angle=0,width=85mm]{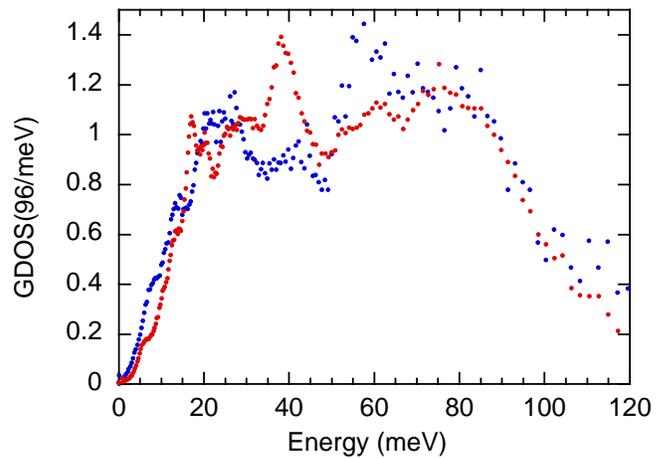}
 \caption{(Color online) Generalized density of states (GDOS) at $T$=~300~K for SCGO (blue) and  SGO (red), obtained using the data shown over a limited $(Q,\omega)$--range in  Fig.~\ref{sqwexp}. 
For SCGO the weight of the paramagnetic contribution has been minimized using only the high $Q$--data ($Q_{el}~>$~3.5\AA$^{-1}$). Due to the remaining magnetic contribution the GDOS at low energy 
shows a faster than Debye growth as compared to SGO. There are distinct differences in the phonon bands in the range 20~meV~$<~ \hbar\omega~<$~70~meV, but we can note  a similar form at low energy, 
the peaked phonon  response seen  in SGO at $\hbar\omega\approx$~5.5~meV and in SCGO at $\hbar\omega\approx$~7~meV (see Fig.~\ref{sqwexp}) produces a characteristic kink at that  same position. A rather 
similar response of the two systems is observed  in the high-energy range dominated by the oxygen modes.}
\label{GDOS1} 
\end{figure}
In addition, the chemical disorder arising from the Cr substitution on the Ga sites, which can in principle be 
modelled, leads in practice to phonon instabilities in the calculations. For these reasons, this study is focused on the non-magnetic material SGO. These results are then used with proper care to explain 
phonon-related phenomena in the magnetic system SCGO. The two systems are isostructural and accordingly  the magnetic Cr sub-lattice sites can be identified crystallographically using the corresponding 
positions of the Ga atoms. Even though it is clear that the phonon response of the two systems cannot be identical both Raman spectroscopy and INS  provide  justification for our approach since the 
spectra of SCGO and SGO display  similar features in the low-lying frequency range. This is reported in Fig.~\ref{raman1} for the Raman data. Both samples show a low-energy mode at $\sim$~53~cm$^{-1}$ 
($\sim$~6.6~meV) and at $\sim$~60~cm$^{-1}$  ($\sim$ 7.4~meV) for SGO and SCGO, respectively. Also the measured neutron density of states of the two systems follow quite similar trends, 
see Fig. \ref{GDOS1}. No attempt was made to eliminate the magnetic contribution that, in spite of being small, influences  the GDOS evaluated for SCGO  in the low-energy range. Supposing that the 
low-energy Raman mode are characteristic of Ga/Cr vibrations we can argue that the shift of positions of the main low-frequency features is partly connected to the different masses of the Cr and Ga atoms, 
52 and 69.72, respectively. The ratio of the observed Raman frequencies ($\sim$0.9) is close to the  square root of the ratio of the atomic masses ($\sim$0.86), as would be expected for a 
vibrational mode in a similar potential involving essentially the Cr or Ga atoms. Comparing the INS response we note a somewhat bigger difference, the peak seen at  5.5~meV  in SGO appears at 7~meV 
in SGCO.  Of course for modes with participation from many atomic species and different interatomic potentials  the simple comparison based on atomic mass only cannot be quantitatively correct.
\begin{figure}
\includegraphics[angle=0,width=85mm]{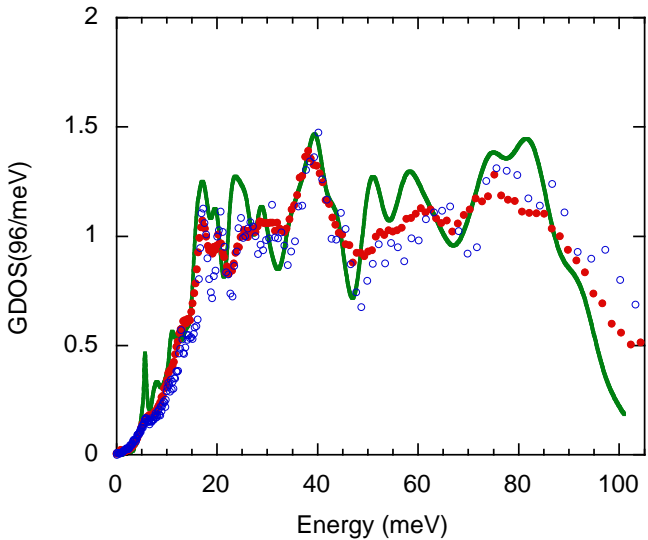}
 \caption{(Color online) Calculated and neutron determined generalized density of states GDOS for SGO show very similar overall features. The results using both IN4 (red dots) and IN6 (blue circles) data are in good agreement  with the calculation (green solid line) and display a similar broadened peak at $\hbar\omega\approx$~5.5~meV, see also Fig.~\ref{lefit}. The calculated data has been convoluted with a gaussian to account for the experimental resolution.}
 \label{dosexpcalc} 
\end{figure}
\begin{figure}
\includegraphics[angle=0,width=70mm]{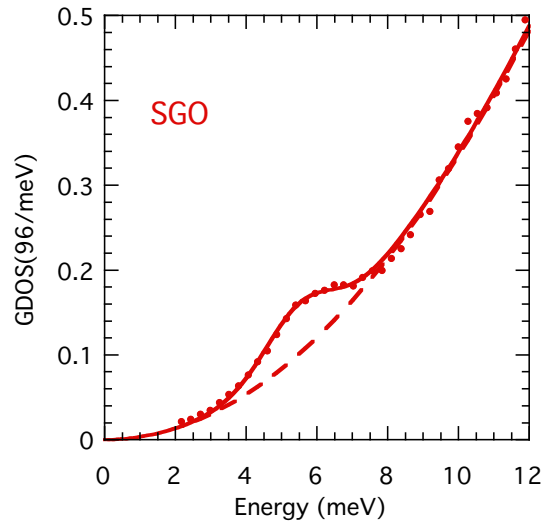}
\caption{(Color online) The low--energy part of the SGO-GDOS can be fitted (full line) with a superposition of a Debye term ($\propto\omega^{2}$, $\hbar\omega_{D}$~=~46~meV) (dashed line) and  single mode peak with a gaussian lineshape. The results using both IN4  and IN6 (not shown)  data are in good agreement and display a similar broadened peak at $\hbar\omega\approx$~5.5~meV, $\delta$~=~2.3~meV (FWHM), indicating that this broadening is not due to instrumental resolution.}
 \label{lefit} 
\end{figure}
\subsection{SGO/SCGO phonon response compared with calculation \label{sgocalc}}
\begin{figure}
\includegraphics[angle=0,width=90mm]{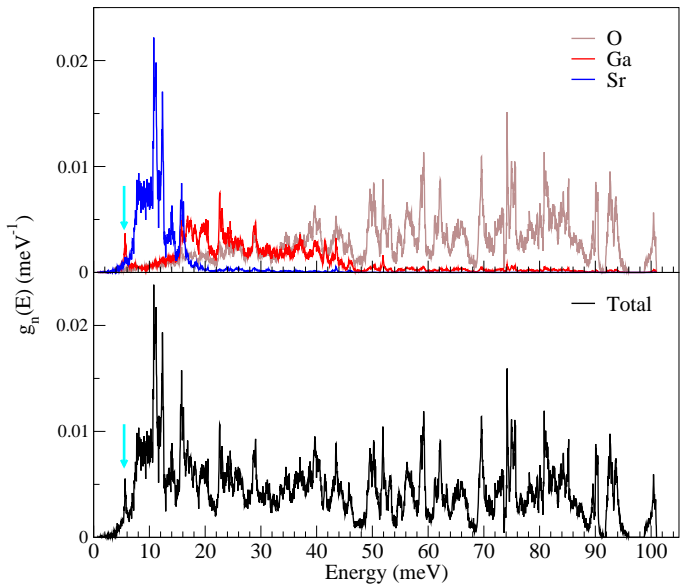}
 \caption{(Color online) Calculated total and partial neutron weighted GDOS for SGO. Arrows indicate the peak in the Ga partial corresponding to the phonon mode of interest at $\hbar\omega$~=~ 5.5~meV.}
 \label{tandpart} 
\end{figure}

\begin{figure}
\includegraphics[angle=0,width=90mm]{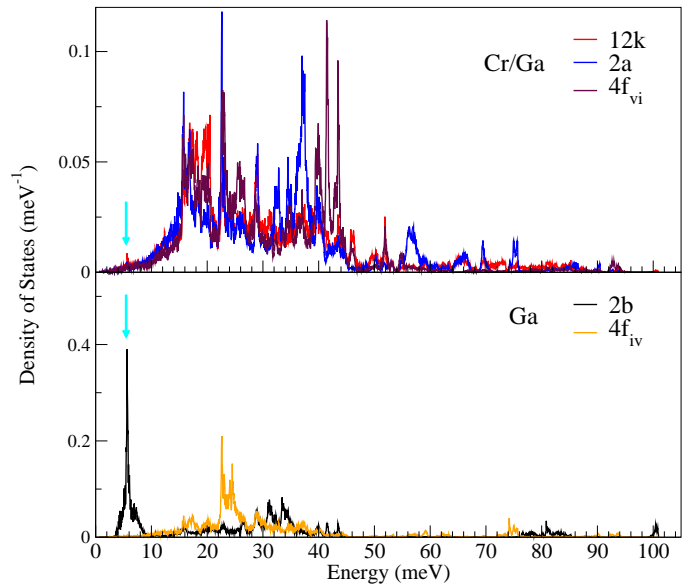}
 \caption{(Color online) \emph{Ab initio} determined partial DOS of Ga-atoms, $g_{Ga,j}$, in SGO. The strongest contributions to the low-energy peak are from the 2b and 12k sites, respectively, as indicated by arrows. See Table~\ref{tab_params} for details of site occupations.}
 \label{gapart} 
\end{figure}

\begin{figure*}
  \centerline{\hbox{ \hspace{-0.2in}
    \epsfxsize=2.7in
    \epsffile{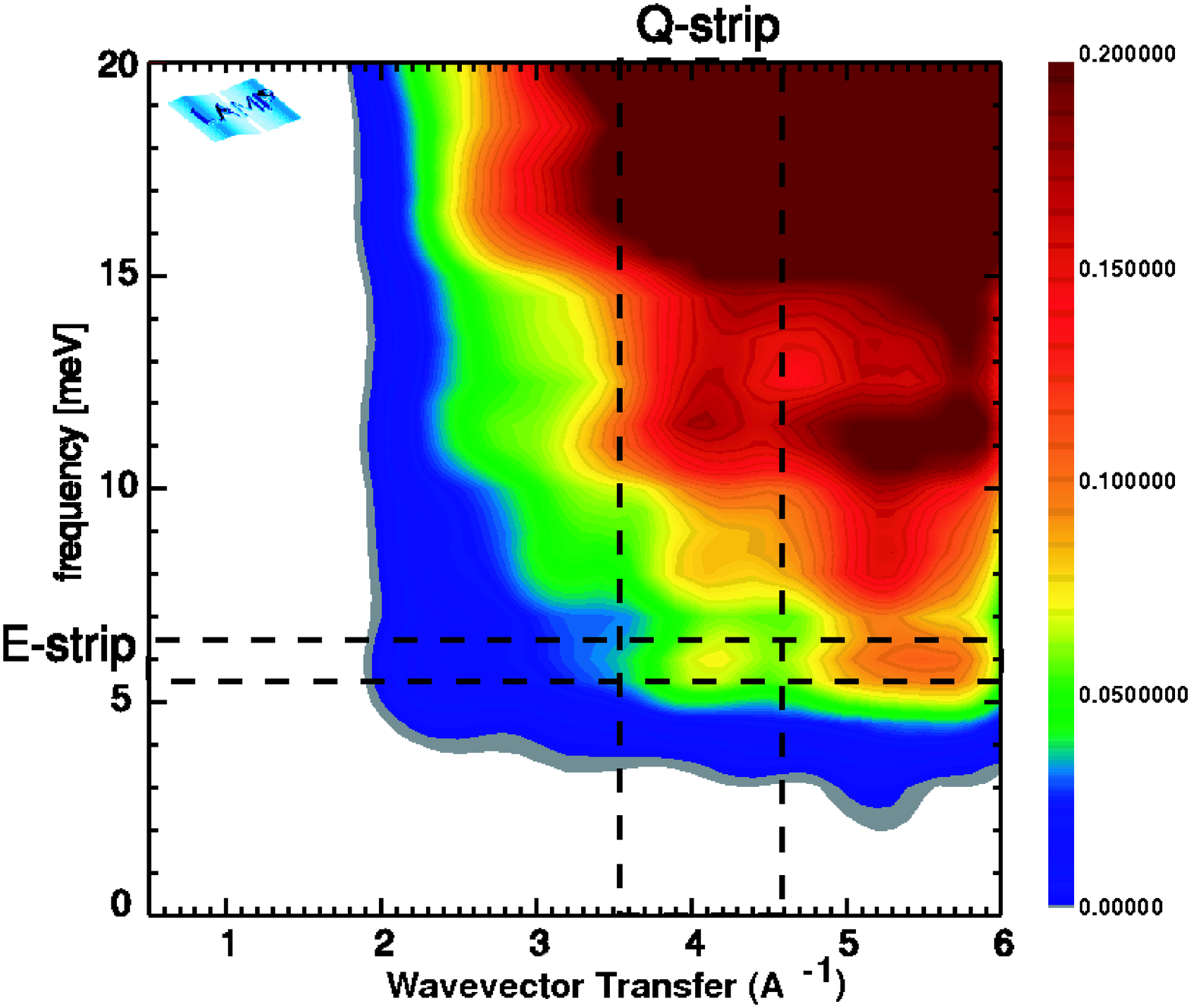}
    \epsfxsize=2.55in
    \epsffile{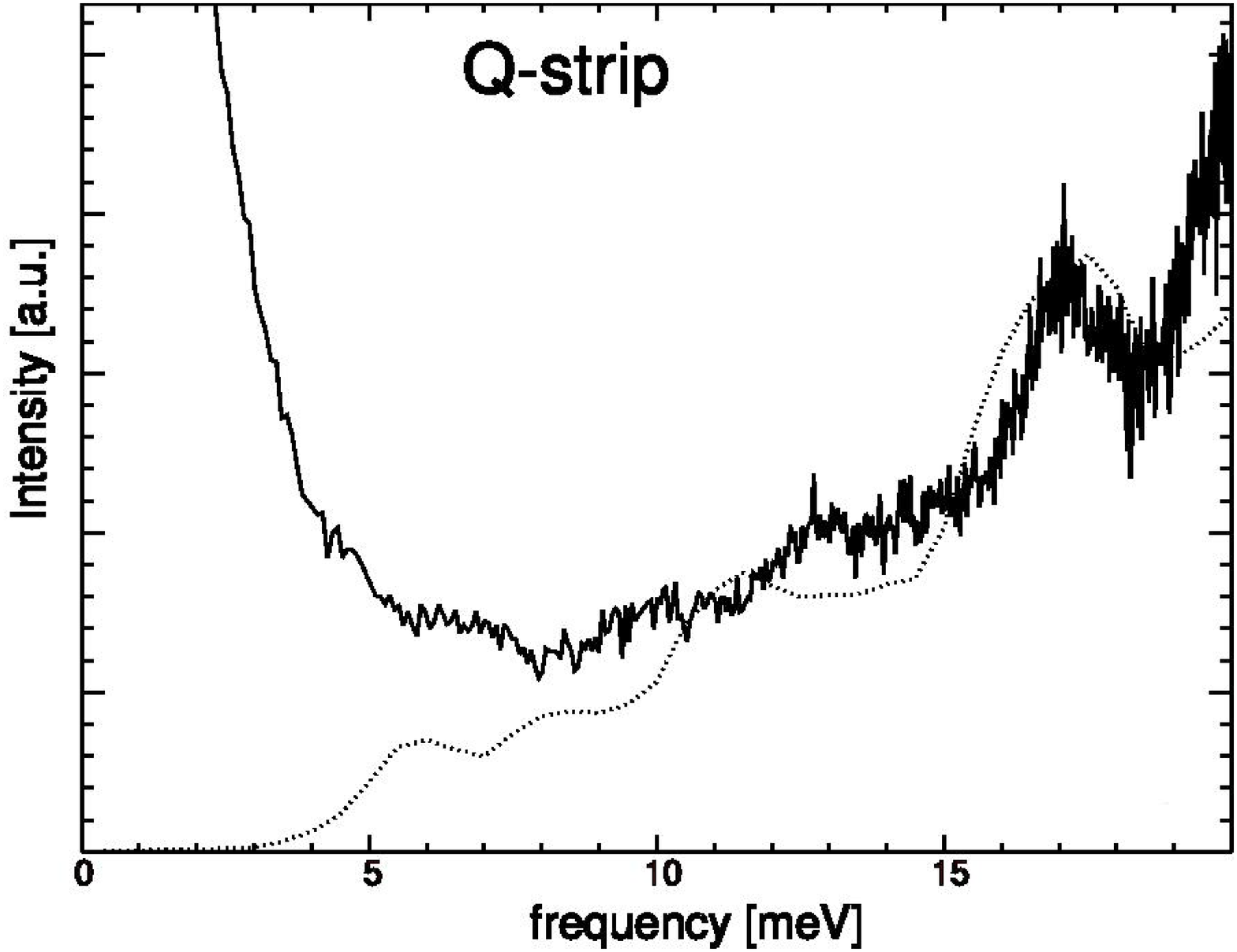}
    \epsfxsize=2.55in
    \epsffile{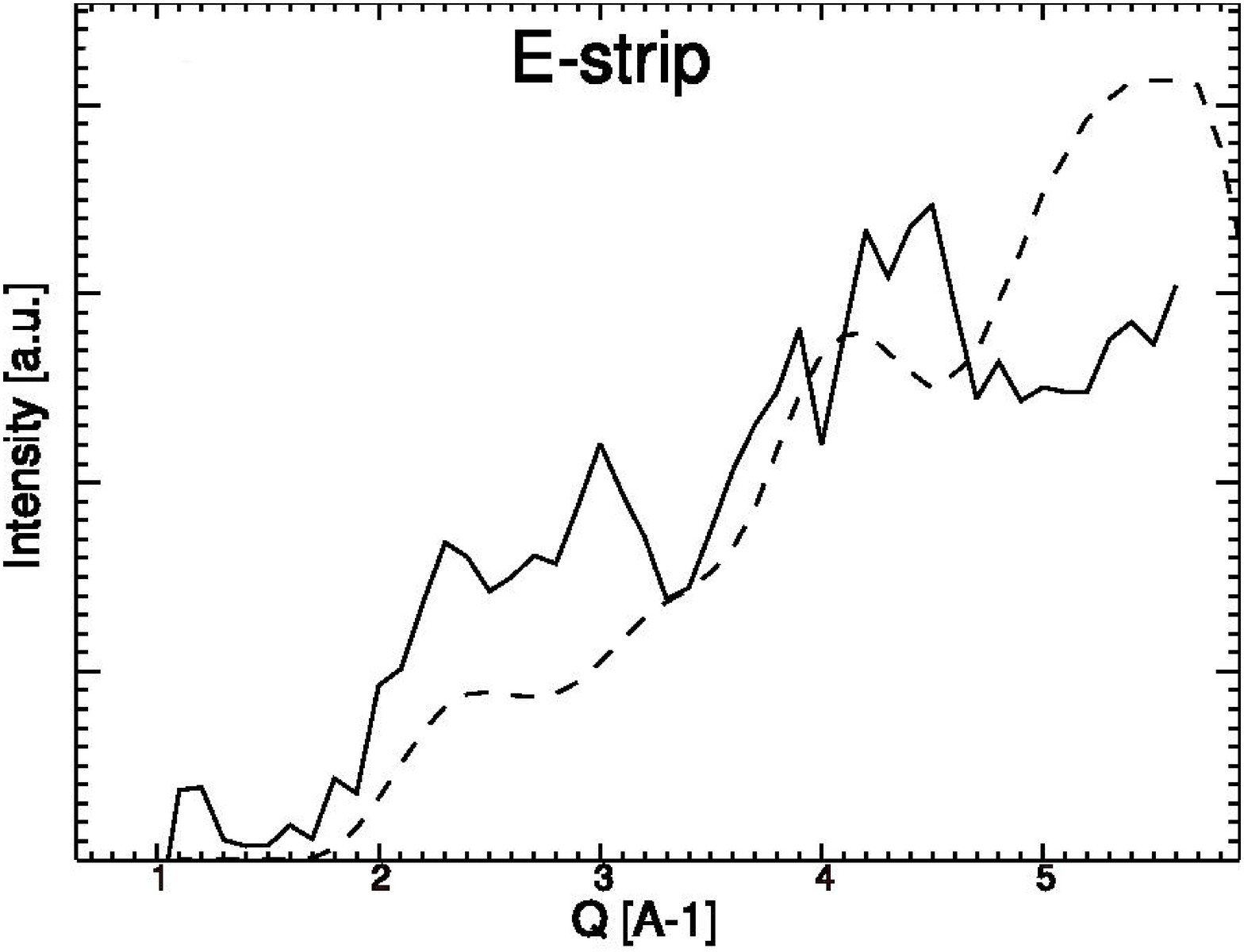}
    }
  }
  \hbox{\hspace{0.6in} (a) \hspace{2.6in} (b) \hspace{2.5in} (c)}
   \caption{(Color online) Simulated $S(Q,\omega)$ map (a) for SGO calculated for the Stokes (one phonon annihilation) process at T=0, Fig. \ref{sqwexp} shows the measured anti-Stokes (phonon creation) 
response weighted by the thermal population. Therefore the relative intensity at higher absolute value of energy transfer appears less in Fig. \ref{sqwexp}. Constant $Q$ cut (b) of measured and 
calculated $S(Q,\omega)$ (solid and dotted lines, respectively). Constant energy cut (c) of measured and calculated $S(Q,\omega)$ ( solid and dashed lines, respectively). The measured cuts shown here are 
the phonon creation response ($\hbar\omega>$~0)  at $T$=2~K, note similarity of the constant energy cut (c) with the one shown in the lower right panel of Fig. \ref{sqwexp} at $T$= 300~K.}
\label{calcmaps}
\end{figure*}

\begin{figure}
\includegraphics[angle=0,width=90mm]{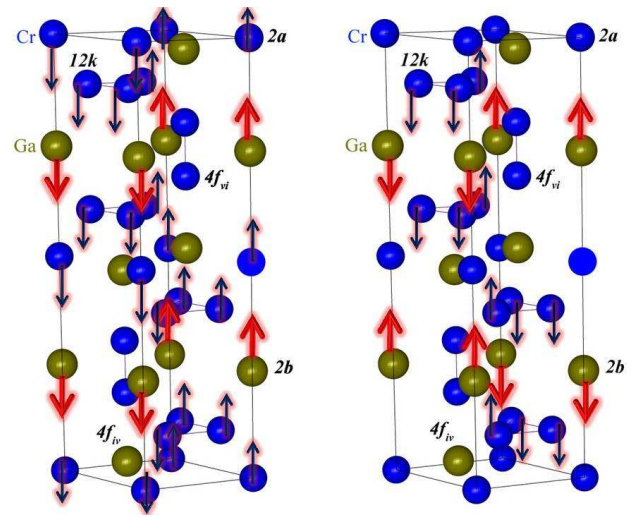}
 \caption{(Color online) Schematic representations of the normal modes in SGO around the M-point at $\sim$ 5.6 meV (left) and at $\sim$ 6.2 meV (right), with arrows indicating the cation displacements. The 2a, 12k and 4f$_{vi}$ are 
sites occupied in major part by Cr in SCGO. The 2b and 4f$_{iv}$ are the sites occupied by Ga only both in SGO and in SCGO.}
 \label{mpointmodes} 
\end{figure}

Fig. \ref{dosexpcalc} compares the \emph{ab-initio} determined and measured GDOS for SGO. In order to compare with experimental data, the calculated GDOS was determined as the sum of the partial 
vibrational densities of states $g_{i}(\omega)$ weighted by the atomic scattering cross sections and masses: GDOS($\omega$)=$\sum_{i} (\sigma_i/M_i)g{_i}(\omega)$, where ($\sigma_i/M_i$ = 0.265 (O), 0.096 (Ga), 0.071 (Sr); i=\{O, Ga, Sr\}). A more detailed look at the neutron data on the SGO  compound is shown in Fig.~\ref{lefit} where one can see that the low-energy phonon response is well described 
by a superposition of a Debye term and a broadened  peak at $\hbar\omega$=5.5 meV. The observed broadening of the peak ($\delta\approx$ 2.3 meV (FWHM) does not depend on the the instrumental resolution 
that is less than the apparent line--width both on IN4 and on IN6. Disregarding the broadening the calculated spectrum is in very good agreement with the observed one, in particular the peak 
at $\sim$ 5.5 meV is very well reproduced and also the relative weight of the peak is in reasonable agreement with calculation. The width of the peak is not resolution limited and we conclude that the 
mode seen in  Raman spectrum reported in Fig.~\ref{raman1} is included in the peak of the GDOS. Due to the magnetic contribution present in SCGO fitting with 
the  the expected $\propto~\omega^{2}$ and a gaussian peak is not applicable. However, mimicking the magnetic contribution with an additional linear term, a reasonable fit can be obtained for the gaussian peak at $\hbar\omega$~= 7~meV  with  a width of $\delta\approx$ 2.4~meV (FWHM) similar to the SGO case (not shown).\\
Based on the above we can consider that  the experimental data validates the computational approach in terms of  the generalized density of states and Raman response. Accordingly we can go further in 
analyzing the calculated partial GDOS. Fig.~\ref{tandpart} shows total (GDOS) and the partials $g_{i}(\omega)$ of SGO. The major contribution of the Ga atoms to the peaked  low-energy response is clear. 
The Ga partial $g_{Ga}(\omega)$ can be further resolved into the contributions from the different crystallographic sites $g_{Ga,j}(\omega)$, see Fig.~\ref{gapart}. This is of interest for the  case of  
SCGO since the magnetic Cr atoms occupy almost fully the sites 2a, 12k and 4f$_{vi}$ that contain only Ga in SGO, see table \ref{tab_params}. It appears that the 2b and 12k site partials, 
$g_{Ga,2b}(\omega)$  and $g_{Ga,12k}(\omega)$ give  the strongest contribution to the total density of states. The 2b site is fully occupied by the non-magnetic Ga also in SCGO while the Cr occupation 
on the 12k and 2a sites, forming the pyrochlore slab, is beyond 90 \% in the least diluted samples~\cite{mond01} (see Table~\ref{tab_params} and Figure~\ref{mpointmodes}). Structural refinements of 
the isostructural magnetoplumbite compounds, including SGO~\cite{grae94}, usually  indicate large Debye--Waller factor for the cation located in the trigonal  bipyramid on the 2b site, and a split 
occupation on the 4e site can be considered (see Table~\ref{tab_params}). \\
The high spectral weight of the $g_{Ga,2b}(\omega)$ partial is consistent with dynamical disorder being the reason for the large and anisotropic thermal displacement  of Ga atom. The 12k and 2a sites constitute the pyrochlore slab magnetic sub-lattice, kagome and triangular planes respectively,  in the SCGO compound, 
and we begin to understand how phonons at 7 meV can drive magnetic relaxation in SCGO.\\
%
\begin{figure}
\includegraphics[angle=0,width=85mm]{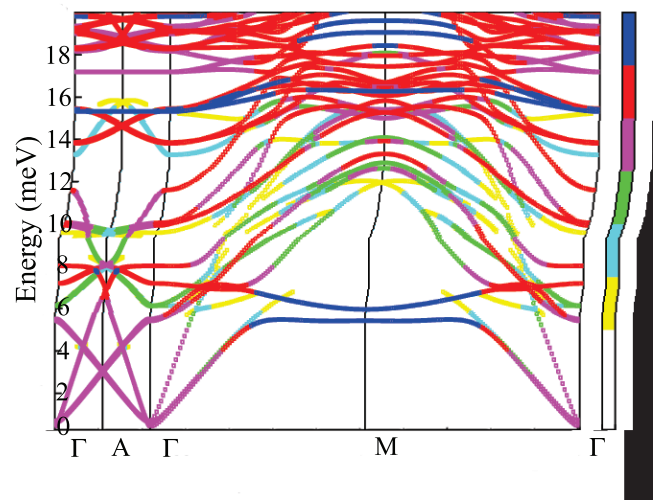}
 \caption{(Color online) Calculated Dispersion relations along the high symmetry directions of the Brillouin zone in SGO, with color coded neutron scattering intensity,  blue color corresponds to the maximum intensity. The Bradley-Cracknell notation is used for the high-symmetry points: $\Gamma$=(0,0,0), A=(0,0,$\xi$), and M=($\xi$,$\xi$,0).}
 \label{dispersion} 
\end{figure}
Calculated phonon dispersion relationships are shown in Fig.~\ref{dispersion} in which  the color-coded intensities correspond to the coherent, dynamic structure factor for one-phonon creation. Close to 
the GDOS low--energy peak ($\sim$ 6 meV) the dispersion curves consists of flat branches with two modes around the M-point having maximal intensity. These modes make the strongest contribution to the DOS 
and we claim we see their signature in form of the gaussian peak seen  in the experimental GDOS, too. At the $\Gamma$-point there are three Raman-active modes close to the GDOS peak but with weaker 
response in the INS. Due to the fact that  the nuclei of the SGO are coherent neutron scatterers the inelastic scattering cross-section has characteristic features even in powder averaged phonon response  
that  can also be resolved in terms of its wave-vector dependence. In order to gain further confidence in the dispersion relations and the overall phonon response, we have calculated the coherent 
structure factor over the whole Brillouin zone. From this data we have generated the $S(Q,\omega)$ map for a powder sample, using the recently developed, powder-averaged lattice dynamics 
approach~\cite{john09, koza08}. The calculated result can be compared with experimental data from IN4, see Fig. \ref{calcmaps}.\\
Comparison by eye of the low-energy spectrum is hampered by the elastic intensity in the experimental data, which is not calculated. Measured and calculated $S(Q,\omega)$ maps are, however, in reasonably good agreement as shown by constant energy and constant $Q$ cuts.\\ 
One can also note that the modulations superposed on the overall $Q^{2}$-dependence of the inelastic signal at energy transfer of 5.5 meV appear periodically at positions close to the successive M-point values at $Q\approx$~1.9 \AA$^{-1}$, 3.1 \AA$^{-1}$, 4.3 \AA$^{-1}$. Note that the measured  data shown in Fig. \ref{calcmaps} is taken at low temperature, $T$~=~2K, and therefore a shorter incident wavelength ($\lambda_{i}$~=~1.8~\AA) was necessary, compared to the setting used to obtain the data shown in Fig.~\ref{sqwexp} ($\lambda_{i}$~=~2.6~ \AA) , in order to reach high enough momentum transfer in the Stokes (neutron energy loss, phonon creation) side of the spectrum.  
The quality of the data at low temperature is worse due to reduced thermal population of phonon states. Also the energy resolution of the instrument is comparably worse. A check of the temperature 
dependence of the phonon response in  SCGO in the vicinity of the spin freezing transition did not indicate any significant  change in the limits of experimental precision. \\
In view of the importance of the two M-point modes (5.6 meV and 6.2 meV), we show schematically their displacement vectors in Fig. \ref{mpointmodes} for the case of the SCGO magnetic compound. Therein the 
displacements of Cr and Ga atoms are highlighted. The sites 12k and 2a belong to the magnetically connected pyrochlore slab network, representing the triangular planes and the kagome planes, 
respectively, and the vibrational characteristics  there should be the most important in the context of the magnetic relaxation. As depicted in Fig.~\ref{mpointmodes} the M-point phonons involve large 
amplitude displacements of the Ga atoms on the 2b sites, and the associated  response of the 2a  and 12 k sites will modulate the magnetic  interactions between these sites. In this context, the observed 
phonon-driven magnetic relaxation could be explained as a consequence of the strong sensivity of the magnetic interactions to subtle variations in the magnetic exchange pathways between adjacent 
Cr magnetic cations. It was shown that the AFM coupling $J$ between nearest-neighbour Cr ions within the pyrochlore slab is very sensitive to the Cr-Cr distance~\cite{limot02}. The J value follows a 
phenomenological law $\Delta J/\Delta d$=450K/\AA. Because phonon calculations could be done for SGO only, it is not possible however to give an accurate value for the J modulation associated with the M 
mode in SCGO. 
\subsection{The case of BSZGO/BSZGCO\label{bsz}}
\begin{figure}
\includegraphics[angle=0,width=85mm]{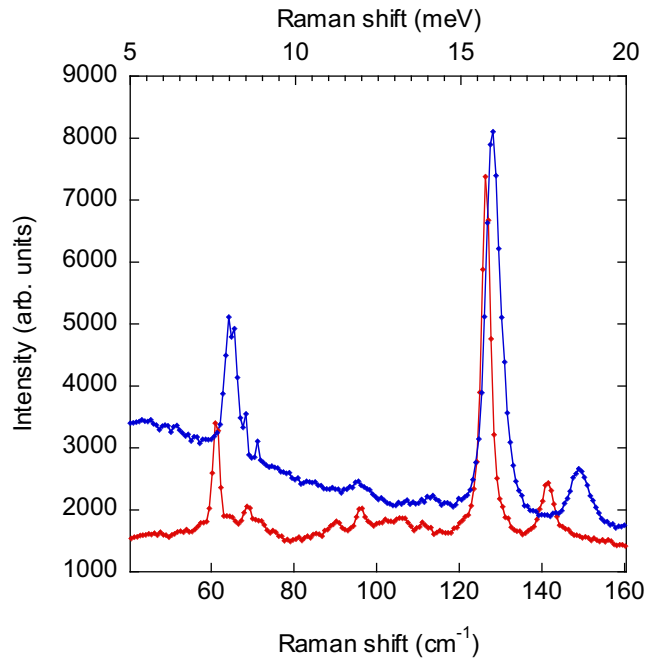}
 \caption{(Color online) Raman spectra on BSZGO (red, lower)  and BSZSCGO (blue, upper), measured at $T$=300~K. Note the low--energy mode similar to the SGO/SCGO case shown in Fig.~\ref{raman1}.}
 \label{bszraman} 
\end{figure}

\begin{figure}
\includegraphics[angle=0,width=85mm]{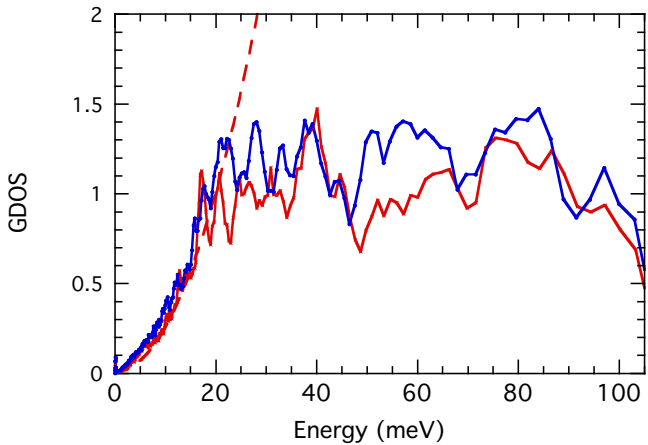}
 \caption{(Color online) GDOS of  BSZGO (blue) compared with that of SGO (red), normalised as explained in the text. For the latter note the Debye like dependence at low temperature (dashed line with $\hbar\omega_{D}$~=~46~meV), see also  Fig.~\ref{lefit}.}
 \label{bszgdos} 
\end{figure}

\begin{figure}
\includegraphics[angle=0,width=85mm]{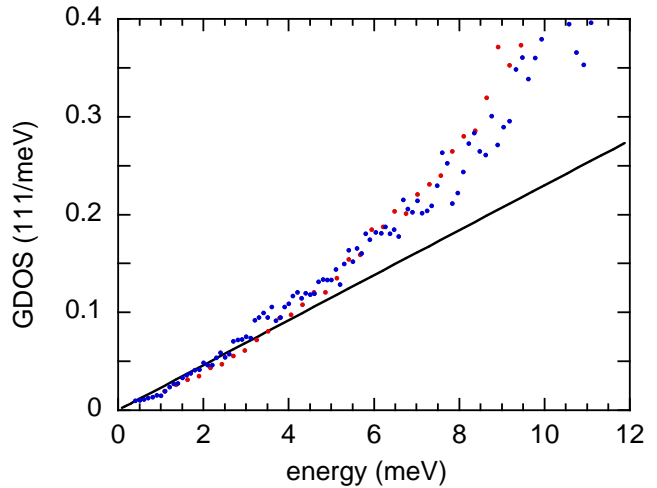}
 \caption{(Color online) The low--energy part of the GDOS of  BSZGO, the solid line  showing the linear non-Debye dependence for $\hbar\omega\to$~0 . Data with  blue and red symbols from IN6 and IN4, respectively. }
 \label{lebsz} 
\end{figure}

We first examine the Raman response, Fig.~\ref{bszraman}. Here again we can see that both the magnetic and non-magnetic compounds display a low-energy Raman 
peak, ressembling the one observed in the SGO/SCGO systems. Also as before, one has a slightly lower energy for the peak in the Cr containing compound, and this situation can be qualitatively ascribed to the difference in atomic mass. 
The experimental GDOS has a  dynamic range comparable to that of SGO/SCGO  and especially the high-energy part shows similar form, Fig.~\ref{bszgdos}. However there is a qualitative difference in the 
experimental GDOS as compared with SGO, the one measured for BSZGO does not show any distinct peak in the range where the Raman peaks are visible. The low-energy phonon response is overall rather 
different, it does not show the characteristic $\omega^{2}$ -dependence typical of acoustic phonon response associated with the Debye-regime. In  fact we observe an almost linear dependence 
GDOS $\propto\omega$ with a progressive upwards curvature that never reaches a quadratic law, see Fig.~\ref{lebsz}. Such a behavior is quite unusual.  For crystalline solids the acoustic phonons with 
linear dispersion should give a quadratic density of states. To explain the unconventional trend we propose that the substitutional disorder in the form  of  50/50 occupation by Zn or Ga on one of the 
2d sites creates a situation that affects the lattice dynamics  and gives rise to a disorder induced dynamic response. We can argue that  the damping of phonon modes due to the lattice disorder leads to 
the observed  non-Debye response of the low--energy modes, not necessarily just the long wavelength acoustic modes but also the ones further out in the Brillouin zone or even at the zone boundary.
\section{Discussion and conclusions\label{disc}}
In the previous section we have provided evidence of the possibility of particular normal modes being responsible for the magnetic relaxation in SCGO. We suggested that  the  activated behavior of the 
relaxation rate is due to the  capacity of the phonons at that energy  to influence  the magnetic system.  In the following we attempt to  examine more generally the role of phonons in the temperature 
dependence of  the magnetic relaxation. We survey a possible minimal model in which the important issue is the capability of the phonon bath to provide energy for the magnetic system. 
\subsection{Magnetic relaxation controlled by  phonon population}
\begin{figure}
\includegraphics[angle=0,width=85mm]{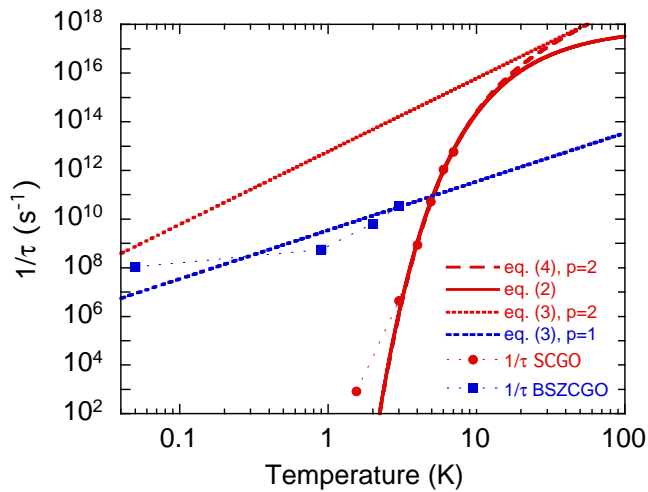}
 \caption{(Color online) Modeling the relaxation, see details in the text. The single frequency activated model of eq.~\ref{sact} follows the experimental data on SCGO over the range 3~K$\lesssim T\lesssim$~7~K  as reported earlier~\cite{mut06} and is in that temperature range indiscernable from the model with cut-off at the same frequency, eq.~\ref{cutoff}. The trend of the  power-law models eq.~\ref{power} does not match the datapoints of neither SCGO (dots) nor BSZCGO (squares). Note that the two models  have been plotted with arbitrarily chosen prefactors.}
 \label{relrate} 
\end{figure}

Let us  assume  that the phonon dominated relaxation rate $\tau^{-1}$ is determined by the number of occupied phonon states, accordingly the temperature dependence is related to the thermal 
population $n_{ph}(T)$. 
 \begin{equation}
\tau(T)^{-1}\propto n_{ph}(T) =  \int \frac{g(\omega)d\omega}{e^{\frac{\hbar\omega}{k_BT}}-1}.
\label{popul}
\end{equation}
In the case of a single mode of energy $\omega_{m}$, we have $g(\omega)\propto \delta(\omega-\omega_{m})$; and the phonon population and the relaxation rate have the activated dependence
\begin{equation}
\tau^{-1}\propto n_{ph}(T) = \frac{1}{e^{\frac{\hbar\omega_m}{k_BT}}-1}\approx e^{-\frac{\hbar\omega_m}{k_BT}}
\label{sact}
 \end{equation}
where the approximation is valid at low temperatures  when $\hbar\omega_{m}/k_{B}T~\gg~1$. This dependence satisfies  the experimental observation concerning the magnetic relaxation in SGCO. However, if 
any low--energy phonon can induce relaxation, one can expect that the low--energy density of states is the important quantity, and then the expected $T$--dependence of the relaxation rate will depend 
on the functional form of the density of states $g(\omega)$. At low temperatures the integration in eq.~\ref{popul} can be taken from zero to infinity and after a transformation of 
variable $x=\hbar\omega/k_{B}T$ one finds that  a power law form $g(\omega)\propto \omega^p$ leads to a power law in $T$,  
\begin{equation}
n_{ph}(T) \propto (k_BT/\hbar)^{p+1} \int _{0}^{\infty}\frac{x^{p}dx}{e^{x}-1}
\label{power}
\end{equation}
where the definite integral can be evaluated  analytically~\cite{bandb06}. Another possible situation might be such that there is a low--energy cut--off $\omega_{c}$ due to e.g. $Q$-- dependence of 
the effective interaction. In this case,  for  $\hbar\omega_{c}/k_{B}T~\gg~1$,   the number of phonons affecting the relaxation rate is evaluated to be
\begin{equation}
n_{ph}(T) \propto (k_BT/\hbar)^{p+1} \int _{\omega_{c}}^{\infty}\frac{x^{p}dx}{e^{x}}\propto(k_BT/\hbar)^{p+1}e^{-\frac{\hbar\omega_c}{k_BT}}
\label{cutoff}
\end{equation}
We have seen above that the experimental GDOS is a superposition of the Debye $\omega^{2}$ and a single mode energy contribution in the case of SGO , while in BSZGO a linear $\omega$--dependence 
appears. In case of independent relaxation processes one can expect that the relaxation rates of each process are additive, 
\begin{equation} 
\tau^{-1}=\sum_{i}\tau_{i}^{-1}
\label{rrsum}
\end{equation}
%
It is straightforward to calculate $n_{ph}(T)$ in the eqs.~\ref{popul}~to~\ref{cutoff}. Note, however, that  to determine a quantitative estimate for the relaxation rate it would be necessary 
to know  the prefactor describing the efficiency of the coupling between the spin system and the phonon bath.  Experimentally we find for the magnetic relaxation in SCGO that 
$\tau^{-1}=\tau_{0}^{-1}\exp(E_{A}/k_{B}T)$ with $\tau_{0}^{-1}~=~7\times10^{17}$~s$^{-1}$. This means that a Debye phonon controlled relaxation of the type $\tau^{-1}=\tau_{D}^{-1}T^{3}$ would dominate at low 
temperatures for values of $\tau_{D}^{-1}\ge10^{12}$~s$^{-1}$, see Fig.~\ref{relrate}, which also shows that both the single mode and the cut-off picture could match equally well the experimental data on 
SCGO in the activated regime.  It is not possible to discern any power law dependence either for SCGO or BSZCGO suggesting that long wavelength acoustic phonons are not effective for  the magnetic 
relaxation process. We can also argue that both in SCGO and in BSZCGO  at the lowest temperatures a cross-over to a quantum  relaxation regime takes place~\cite{mut06,mut07}, and we know that at higher 
temperatures a  quasi-elastic response with a width proportional to temperature prevails~\cite{mond00}. Meanwhile, it is clear that the  limited temperature range of the observations does not allow a full 
analysis of the  relaxational dynamics. We suggest that the most effective modes are the ones in the vicinity of the $M$--point of the reciprocal lattice, indeed these zone boundary modes have a spatial 
structure on  the scale of the in plane lattice parameter, and therefore match the nearest neighbor distance between the localized magnetic moments.  With respect to the situation with BSZCGO, we have 
observed no specific activation energy and it is not possible to see any well-defined power--law dependence in the temperature dependence of the relaxation rate. We evoked earlier (Sect. \ref{bsz}) the 
possibility of the disorder induced modification of the lattice dynamics being the origin of the faster relaxation, due to heavily damped phonon modes. Note that here again the space scale is typical of 
the pyrochlore slab in--plane unit due to the half/half Zn/Ga substitutional disorder. It is conceivable that this disorder has a strong influence especially on the $M$-point zone boundary modes  that, 
in case of an overdamped response, can achieve a  quasi-elastic character, and give rise to the linear GDOS as observed, as well as to the faster relaxation seen in BSZCGO. Anyhow it is clear that 
the $T^{2}$ dependence that one might expect with reference to the model calculation  is not observed. 
\subsection{ Other aspects of the relaxation mechanism}
Our results suggest an important role of lattice vibrations in the low-temperature spin dynamics of the frustrated magnets SCGO and BSZCGO. However, one cannot claim that spin--phonon coupling alone could 
be the origin of  the complex relaxational behavior. The systems studied are disordered due to the non-magnetic dilution that has been widely studied  and is a known factor in the pyrochlore slab 
compounds~\cite{mond01,limot02,bono05}. It is even surprising that even though the phonon properties of the SGO/SCGO lattices appear somewhat better defined when 
compared to the BSZGO, it is in SCGO that very strongly stretched time decay occurs.  A complete microscopic picture of the relaxation process is still to be constructed and one can imagine that the fine details of the magnetic states of 
lowest energy are of major importance in such a pursuit. Pursuing  the analysis  might be extremely difficult for systems like the pyrochlore slabs but in simpler cases details of spin-lattice coupling 
have been already examined and could pave the way for further work~\cite{sush05, fennie06}. In the general context of frustrated magnetism our results point out a trend that has been already evoked, the 
low--temperature and ground--state properties are highly sensitive to system dependent perturbations. Nevertheless  the present work pinpoints  the importance of lattice vibrations and their microscopic 
character in the control of the low--energy spin dynamics. 
\section{Summary}
We have presented a detailed investigation of the phonon spectra  in the pyrochlore slab compounds SrCr$_{9x}$Ga$_{12-9x}$O$_{19}$ (SCGO) and  Ba$_{2}$Sn$_{2}$ZnCr$_{7x}$Ga$_{10-7x}$O$_{22}$ (BSZCGO) based 
on \emph{ab-initio} lattice dynamics simulations, inelastic neutron scattering and Raman measurements, in order to investigate the origin and mechanisms of the recently observed~\cite{mut06} phonon-driven 
magnetic relaxation in SCGO. Since the magnetic signal dominates the neutron scattering response at low energy transfer, we have performed new experiments on the isostructural non-magnetic material 
SrGa$_{12}$O$_{19}$ (SGO). Moreover,  the chemical disorder is difficult to include in the ab-initio lattice dynamics calculations so these have also focussed on SGO. Results of the calculations for 
the GDOS and $S(Q,\omega)$  for SGO are in good agreement with the experimental data.  Neutron and Raman experiments show the similarity of the phonon response in the magnetic system with respect to the 
non-magnetic counterpart. The calculated  partial density of states for SGO indicates that the strongest contribution to the total density of states stems  from the vibrations of the Ga--atoms on the 2b 
and 12k sites of the lattice, associated with flat dispersion branches centered at the zone boundary M--point in the $k$-space. Thus we conclude that the phonons response in SCGO corresponding to that at 
the M--point in SGO comprises  the partial contribution of the atoms residing on the 12k sites of the magnetic sub-lattice of  SCGO.  The most relevant M-point modes have displacement vectors that modulate 
the distances between Cr sites in the kagome layers of the pyrochlore slab and therefore can effectively couple with the magnetic moments. The activation energy for magnetic relaxation is equal to the 
position of the characteristic peak in the experimental GDOS. Calculated energies of two of the Raman active $\Gamma$ -point modes are included in the energy range of this peak but they  appear with less 
weight  in the calculated GDOS.\\
In comparison BSZCGO, the other pyrochlore slab antiferromagnet shows a qualitatively different low--energy phonon response with an enhanced non-Debye low--energy response. We suggest that the faster 
relaxation without a particular energy scale is a consequence of this circumstance, which we attribute to the particular Ga/Zn disorder present in this compound. Future computational and theoretical work 
is called for to examine  the interplay of magnetic interactions and the microscopic mechanisms associated with  specific phonon displacements for better understanding of the reported phenomena.
\begin{acknowledgments}
C.P. thanks J.Y. Mevellec for the Raman measurements.
\end{acknowledgments}
\end{document}